\documentclass[twocolumn,aps,prb,amsmath,floatfix]{revtex4}
\usepackage{graphicx}
\begin{document}
\title{Coulomb Correlations and Orbital Polarization 
       in the Metal Insulator Transition of VO$_2$}
\author{A. Liebsch$^1$, H. Ishida$^2$, and G. Bihlmayer$^1$}
\affiliation{
\mbox{
$^1$Institut f\"ur Festk\"orperforschung,~Forschungszentrum J\"ulich,
        ~52425 J\"ulich, Germany} \\
\mbox{
$^2$College of Humanities and Sciences, Nihon University,~Sakura-josui,
        ~Tokyo 156, Japan}  \\
}
\begin{abstract}
The quasi-particle spectra in the metallic rutile and insulating 
monoclinic phases of VO$_2$ are shown to be dominated by local Coulomb 
interactions. In the rutile phase the small orbital polarization among 
V $3d$ $t_{2g}$ states leads to weak static but strong dynamical 
correlations. In the monoclinic phase the large $3d$ orbital polarization 
caused by the V--V Peierls distortion gives rise to strong static 
correlations which are shown to be the primary cause of the insulating 
behavior. 
\\  \\
PACS numbers: 71.20.Be, 71.27.+a. 79.60.Bm
\end{abstract}
\maketitle

\section{Introduction}

The metal insulator transition in VO$_2$ has been intensively studied
for a long time. At 340 K the resistivity changes by several orders 
of magnitude.\cite{morin} The high-temperature metallic phase has a 
rutile structure, while the low-temperature insulating phase is 
monoclinic ($M_1$), with a zigzag-type pairing of V atoms along the 
$c$ axis. Both phases are non-magnetic. Although this transition 
is widely regarded as a Mott-Hubbard transition, 
\cite{pouget,zilbersztejn,mott,rice}
the role of the Peierls distortion of the crystal structure in the 
insulating phase has been the topic of intense debate.\cite{goodenough,%
caruthers,wentzcovitch,eyert,huang,korotin,continenza,laad} 
The discovery of a second insulating phase ($M_2$),
\cite{pouget} in which only half of the V atoms form pairs while the
other evenly spaced chains behave as magnetic insulators, suggested 
that both low temperature phases should be regarded as Mott-Hubbard 
and not as Peierls band insulators.\cite{zilbersztejn,rice}  
Although the role of the Coulomb interaction in the metal insulator 
transition of VO$_2$ has been studied previously, 
\cite{huang,korotin,continenza,laad} a consistent description of 
the rutile and monoclinic phases is not yet available.
  
The aim of the present work is to elucidate the interplay of Coulomb 
correlations and orbital polarization in the quasi-particle spectra 
of VO$_2$. Analyzing recent photoemission data\cite{okazaki,tjeng} 
we demonstrate that the metallic and insulating phases show evidence 
of strong local interactions which manifest themselves in distinctly 
different ways because of the different degree of orbital polarization 
in the rutile and monoclinic structures. The key aspect of the metallic 
phase is the small orbital polarization among V $3d$ $t_{2g}$ states, 
implying weak static but strong dynamical correlations. Thus, the 
spectra reveal band narrowing and reduced weight of the coherent peak 
near the Fermi level, and an incoherent satellite feature associated 
with the lower Hubbard band. The monoclinic phase, in contrast, is 
characterized by a pronounced $t_{2g}$ orbital polarization induced 
by the V--V Peierls distortion. The local Coulomb interaction 
therefore leads to strong static correlations and to the opening of 
an excitation gap. The influence of non-diagonal coupling
among $t_{2g}$ orbitals on the size of the gap is investigated 
within the static limit and is found to be small. 

In the following sections we discuss results obtained within several 
theoretical approaches, such as the dynamical mean field theory 
(DMFT)\cite{DMFT}, the local density
approximation plus Hubbard $U$ (LDA+U)\cite{LDA+U} and $GW$\cite{GW} 
methods. We argue that none is presently capable of explaining all 
of the observed phenomena in a consistent manner. Instead, we 
focus on the merits and failures of these approaches with the aim 
of highlighting the different roles of Coulomb correlations and 
orbital polarization in the metallic and insulating phases of VO$_2$.

\section{Results and Discussion}

\begin{figure}[t!]
  \begin{center}
  \includegraphics[width=4.5cm,height=7cm,angle=-90]{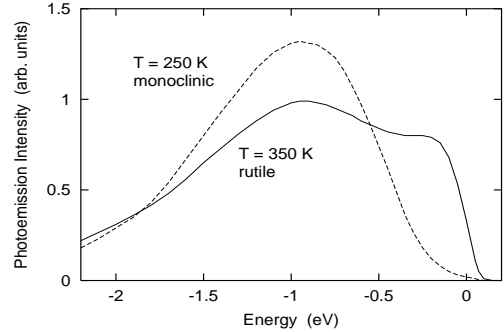}
  \end{center}
  \vskip-3mm
\caption{
Measured photoemission spectra for VO$_2$ films on TiO$_2$ in the V $3d$ 
band region after background subtraction. Solid curve: $T = 350$\,K 
(metallic rutile phase); dashed curve: $T = 250$\,K (insulating monoclinic 
phase); $E_F=0$. \cite{okazaki}
}\end{figure}
       
Fig.~1 shows photoemission spectra for VO$_2$ films (150 \AA\ thick) 
grown on TiO$_2$(001) at binding energies corresponding to the 
V $3d$ bands (21.2~eV photon energy).\cite{okazaki} Emission from 
O $2p$ states extends from $-2$\,eV to about $-8$\,eV. As a result of 
compressive stress from the substrate \cite{muraoka} 
the transition temperature is lowered from 340~K to 290~K. The high 
temperature spectrum shows a Fermi edge characteristic of metallic 
behavior which is absent in the low temperature spectrum. While this 
trend is in agreement 
with previous photoemission measurements,\cite{PES,shin} the data in Fig.~1 
reveal two spectral features in the metallic phase: a peak close to $E_F$ 
and a second one near $-1$\,eV. In the insulating phase, the peak near 
$E_F$ disappears and the feature near $-1$\,eV becomes more intense. Recent 
VO$_2$ photoemission spectra taken at 520 eV photon energy \cite{tjeng} 
agree with the ones shown in Fig.~1, except for a considerably greater 
relative weight of the coherent peak near $E_F$ in the metallic phase.   
These spectral changes are consistent with the observation that satellite
peaks in low photon energy spectra tend to be more pronounced as a result
of a surface induced enhancement of Coulomb correlations. 
\cite{maiti,sekiyama,liebsch03} 

Before analyzing the photoemission data we discuss the 
single particle properties of VO$_2$ obtained using density functional 
theory. We have carried out full potential linearized augmented plane wave 
(LAPW) calculations for the rutile and monoclinic structures using the 
experimental lattice parameters and treating exchange correlation within 
the generalized gradient approximation (GGA).\cite{perdew} 
Due to the octahedral crystal field, the states near $E_F$    
have V $3d$ $t_{2g}$ character. They are separated by a 
small gap from the empty $e_{g}$ states, and from the O $2p$ 
states by a gap of about 1.0~eV. The occupancy of the $t_{2g}$ manifold 
is $3d^1$. Our results qualitatively confirm previous LDA calculations.
\cite{wentzcovitch,eyert,huang,korotin} 
Fig.~2(a) shows a comparison of the V total $t_{2g}$ density of states 
for the rutile and monoclinic phases of VO$_2$. Although there are         
differences in detail, the overall width of these distributions and 
the shape of the occupied region are similar for both structures. 
Evidently, the GGA/LDA does not predict the insulating nature 
of the monoclinic phase. Moreover, on the basis of these densities
one would not expect correlations to play very different roles in 
the two phases. However, if we plot the subband contributions to the 
$t_{2g}$ density, the two structures are very different, as shown in 
Figs.~2(b) and (c). Whereas in the rutile phase the $t_{2g}$ bands 
have similar occupation numbers, 
in the monoclinic phase the $d_{x^2-y^2}$ band is significantly more 
occupied than the $d_{xz,yz}$  bands.
(We adopt the local coordinate system of Ref.~\onlinecite{goodenough}, i.e., 
$x$ denotes the $c$ axis, while $y$ and $z$ point along the diagonals 
of the $a,b$ plane.) The origin of this orbital polarization is the 
V--V dimerization along the $c$ axis. 
The $d_{x^2-y^2}$ band splits into bonding and anti-bonding components, 
while the $d_{xz,yz}$ bands are pushed upwards due to shortening of V-O 
distances. Investigation of the energy bands shows that, in agreement 
with earlier work,\cite{wentzcovitch,eyert} the top of the bonding 
$d_{x^2-y^2}$ bands is separated by a slight negative gap from the 
bottom of the $d_{xz,yz}$ bands. In the following subsection we show
that different degree of orbital polarization in the rutile and 
monoclinic phases has a pronounced effect on the quasi-particle 
spectra of VO$_2$.

\begin{figure}[t!]
  \begin{center}
  \includegraphics[width=4.5cm,height=7cm,angle=-90]{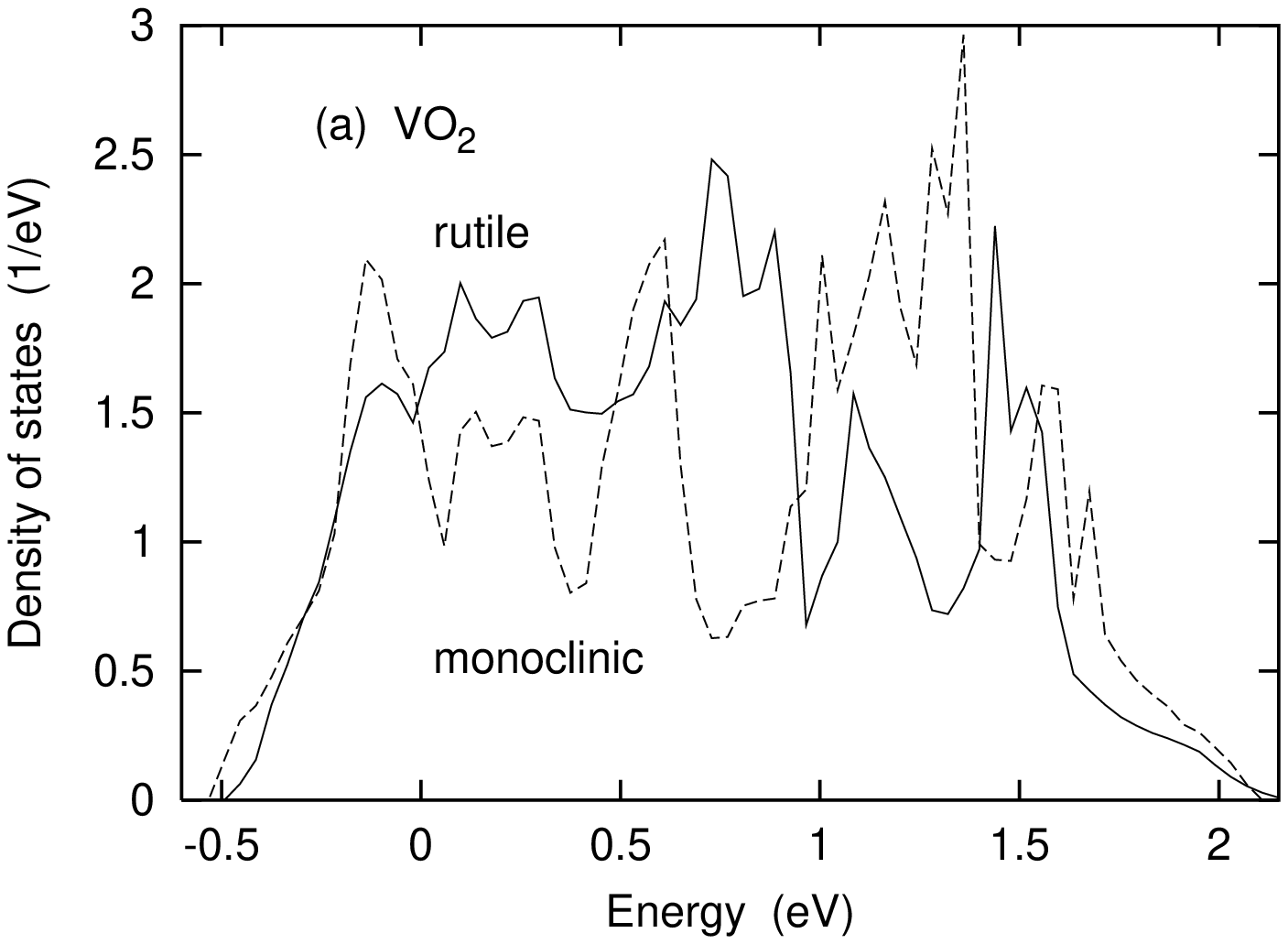}
  \includegraphics[width=4.5cm,height=7cm,angle=-90]{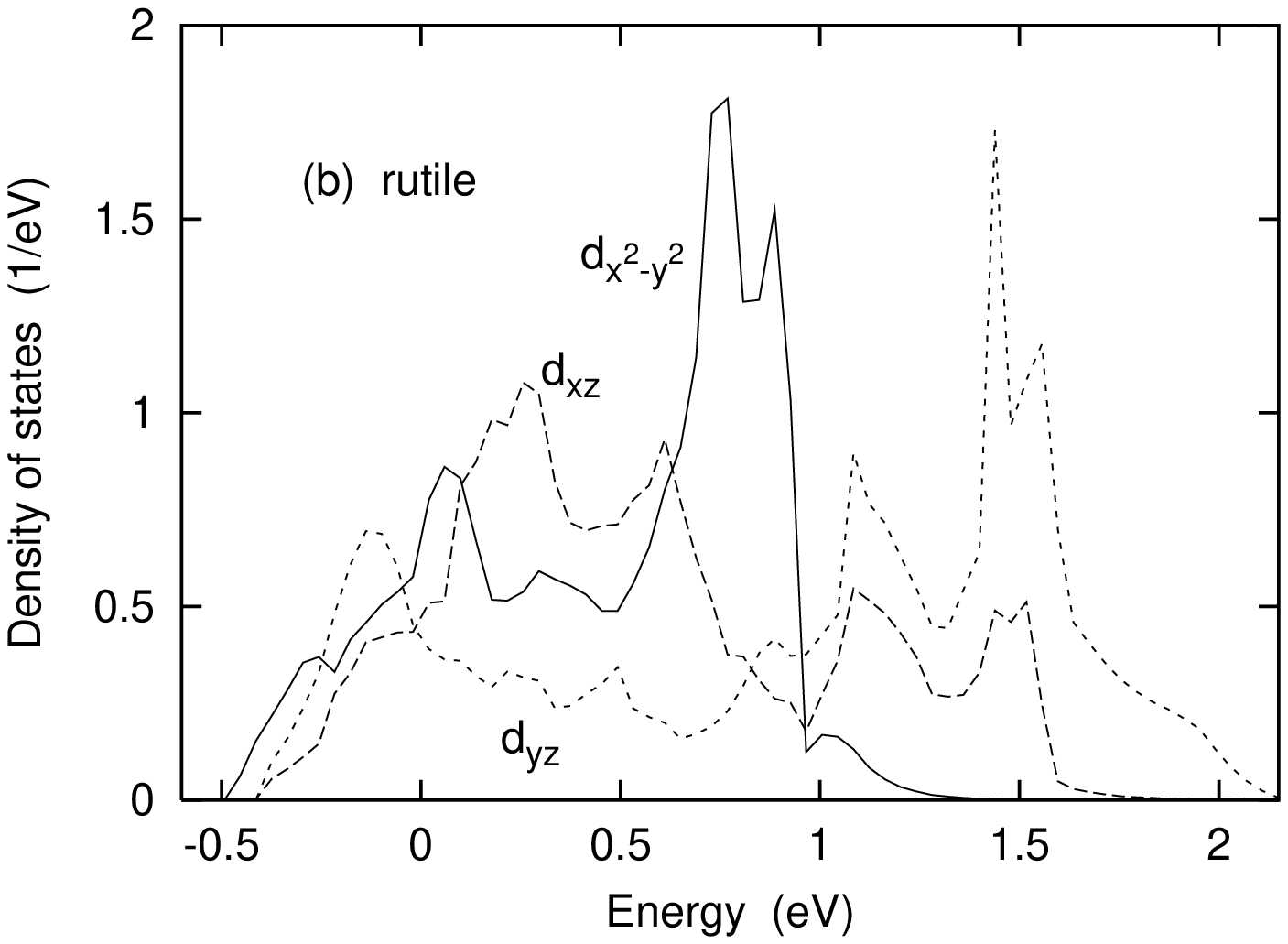}
  \includegraphics[width=4.5cm,height=7cm,angle=-90]{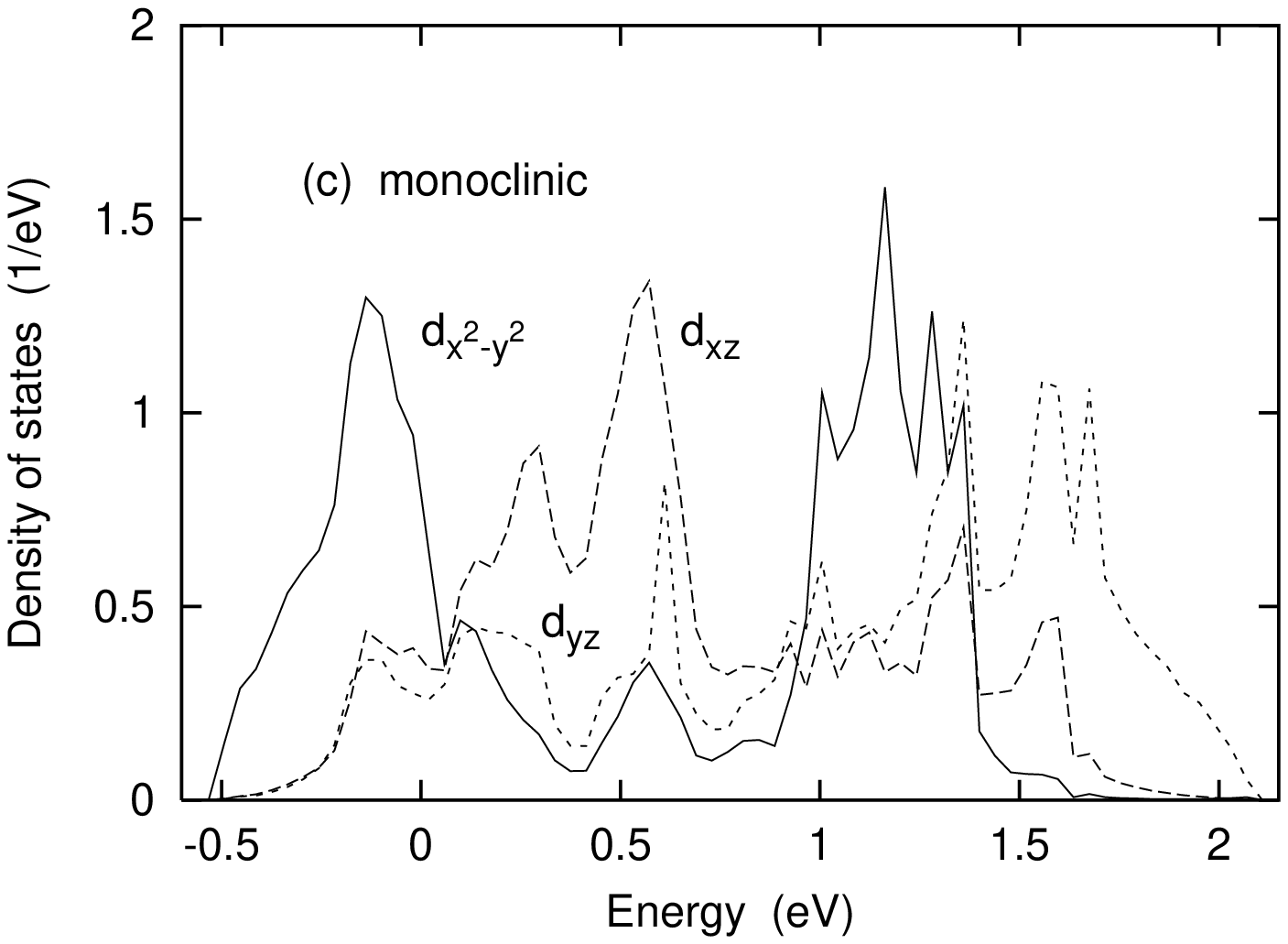}
  \end{center}
  \vskip-4mm
\caption{
VO$_2$ $3d$ density of states calculated within LAPW method.
(a) Total $t_{2g}$ densities for rutile and monoclinic phases; 
(b) and (c) $t_{2g}$ density of states components for rutile and 
monoclinic phases; $E_F=0$.
}\end{figure}

\subsection{Metallic Rutile Phase}

Let us first discuss the rutile phase. Comparing the $t_{2g}$ density 
of states with the photoemission spectra it is plausible to associate 
the feature close to $E_F$ with emission from metallic V $3d$ states. 
The peak near $-1$\,eV, however, lies in the gap between V $3d$ and O 
$2p$ states and cannot be understood within the single particle picture. 
To describe the spectra in the rutile phase it is clearly necessary to 
account for dynamical Coulomb correlations. For the evaluation of the 
quasi-particle distributions we use the Dynamical Mean Field Theory 
combined with the multiband Quantum Monte Carlo (QMC) method.
\cite{DMFT} Since hybridization among $t_{2g}$ states is weak the
local self-energy is taken as diagonal in orbital space. The $t_{2g}$ 
density of states components shown in Fig.~2(b) then serve as input 
quantities accounting for the single particle properties of the rutile 
structure. The local Coulomb interaction defining the quantum impurity 
problem is characterized by intra- and inter-orbital matrix elements $U$ 
and $U'=U-2J$, where $J$ is the  Hund's rule exchange integral. 
According to constrained LDA calculations, $U\approx4.2$\,eV and 
$J\approx0.8$\,eV.\cite{korotin,UJ} The calculations are performed 
at $T\approx500 $~K with up to $10^6$ sweeps. The quasi-particle 
distributions are obtained via maximum entropy reconstruction. 
\cite{jarrell}

Fig.~3(a) shows that, in contrast to the single particle density of
states, the calculated $t_{2g}$ quasi-particle spectra for the rutile 
phase of VO$_2$ exhibit two spectral features, a coherent peak near 
$E_F$ and a lower Hubbard band near $-1$\,eV, in agreement with 
experiment. The peak near $E_F$ accounts for the band narrowing and 
lifetime broadening of the metallic states whereas the Hubbard band
is associated with incoherent emission. Since the $t_{2g}$ subbands 
have comparable single particle distributions their quasi-particle 
spectra reveal similar correlation features. Moreover, because of the 
weak orbital polarization in the rutile structure static correlations 
are negligible. Thus, in the metallic phase the spectral weight 
transfer between coherent and incoherent contributions to the 
spectrum is primarily the result of dynamical correlations.  
    
We point out that, in view of the approximate nature of the model 
underlying the DMFT, quantitative agreement with photoemission data 
cannot be expected. On the theoretical side, the consideration of 
purely on-site Coulomb interactions and the neglect of the momentum 
variation of the self-energy permit only a qualitative analysis of 
the spectra. In addition, there exists some uncertainty regarding 
the precise values of the Coulomb and exchange energies. Finally, the 
DMFT results depend on the temperature used in the QMC calculation. 
The comparison with results obtained for slightly different values of 
$U$, $J$ and $T$, however, gives us confidence that in the metallic 
phase the transfer of spectral weight from the coherent peak near 
$E_F$ to the satellite region near $-1$~eV is qualitatively reliable 
and consistent with analogous dynamical correlation effects in other 
$3d^1$ transition metal oxides, such as SrVO$_3$. 
\cite{sekiyama,liebsch03,pavarini}
As we discuss below, the local DMFT treatment predicts the monoclinic 
phase to be also metallic. The metal insulator
transition in VO$_2$ is therefore not achieved simply by lowering the
temperature in the rutile phase. The lattice transformation from 
the rutile to monoclinic structure must be taken into account. 
Therefore, the DMFT results shown in Fig.~3(a) for the rutile 
structure at $T=500$~K can be regarded as representative of 
correlation induced behavior in the metallic domain. 

On the experimental side, as pointed out above, it is important to 
recall that photoemission data taken at low photon energies represent 
a superposition of bulk and surface contributions. Since correlation 
effects are observed to be more enhanced at surfaces, 
\cite{maiti,sekiyama,liebsch03} the relative intensity of the 
satellite peak near $-1$\,eV in the 21.2~eV spectra shown in Fig.~1 
is considerably more pronounced than at high photon energies 
\cite{tjeng} at which primarily bulk-like valence states are detected.  

Dynamical effects in the metallic phase may also be evaluated within 
the $GW$ approach \cite{GW} which treats long range Coulomb interactions 
in the random phase approximation (RPA) and which has proven rather
useful to describe excitation spectra of weakly correlated systems.
\cite{GWreview} 
Because of the neglect of multiple electron-electron and hole-hole 
scattering processes, this scheme fails when local Coulomb interactions 
are important.\cite{gunnarsson} Presumably, therefore, for VO$_2$ the 
GW method does not reproduce the lower Hubbard band. The satellite in 
the rutile phase is, of course, also beyond the scope of the static 
LDA+U approach.\cite{comment} The qualitative agreement between the 
measured high-temperature spectra and the theoretical results shown 
in Fig.~2(a) suggests that the DMFT captures the key spectral weight 
rearrangement induced by dynamical correlations.

\begin{figure}[t!]
  \begin{center}
  \includegraphics[width=4.5cm,height=7cm,angle=-90]{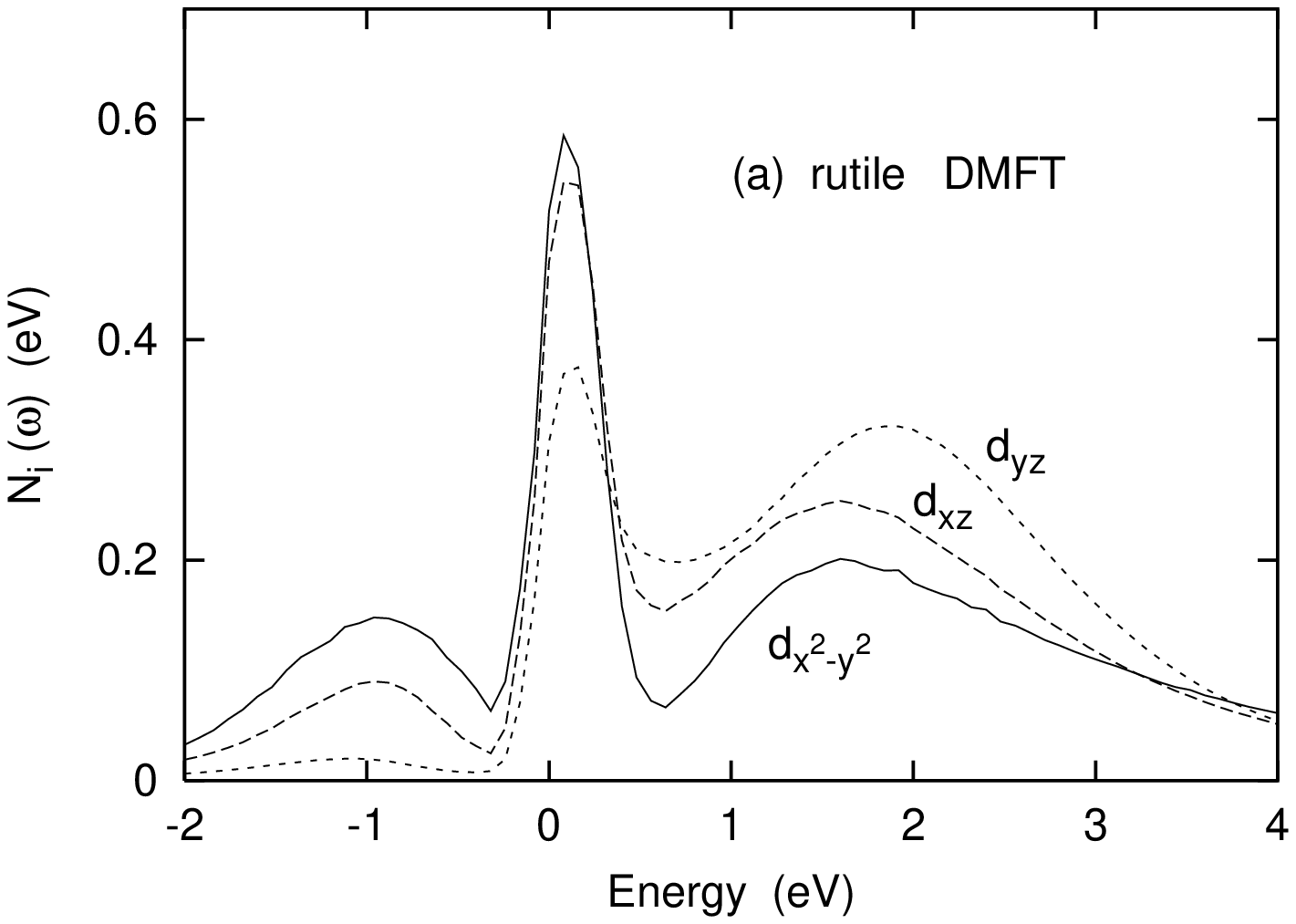}
  \includegraphics[width=4.5cm,height=7cm,angle=-90]{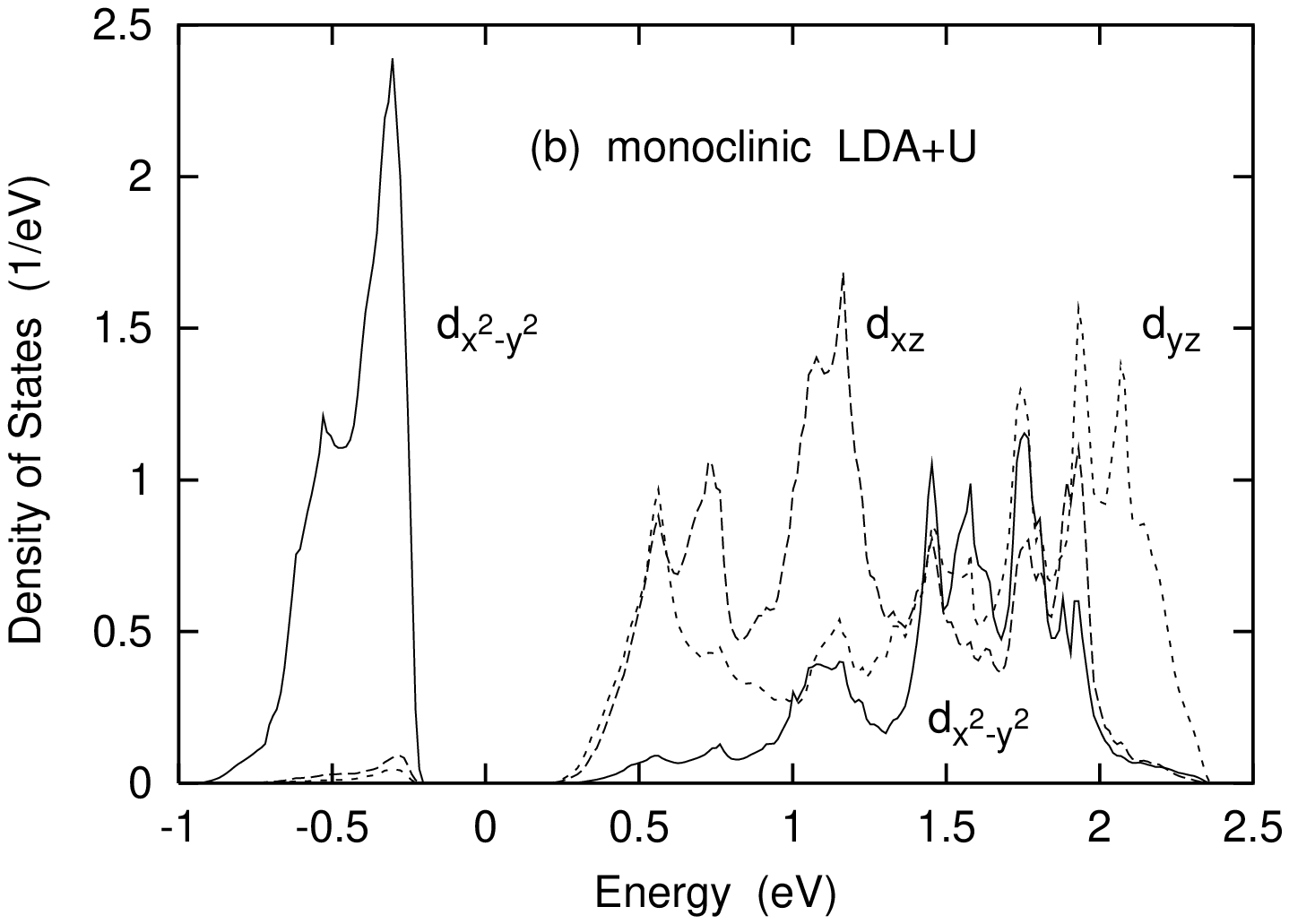}
  \includegraphics[width=4.5cm,height=7cm,angle=-90]{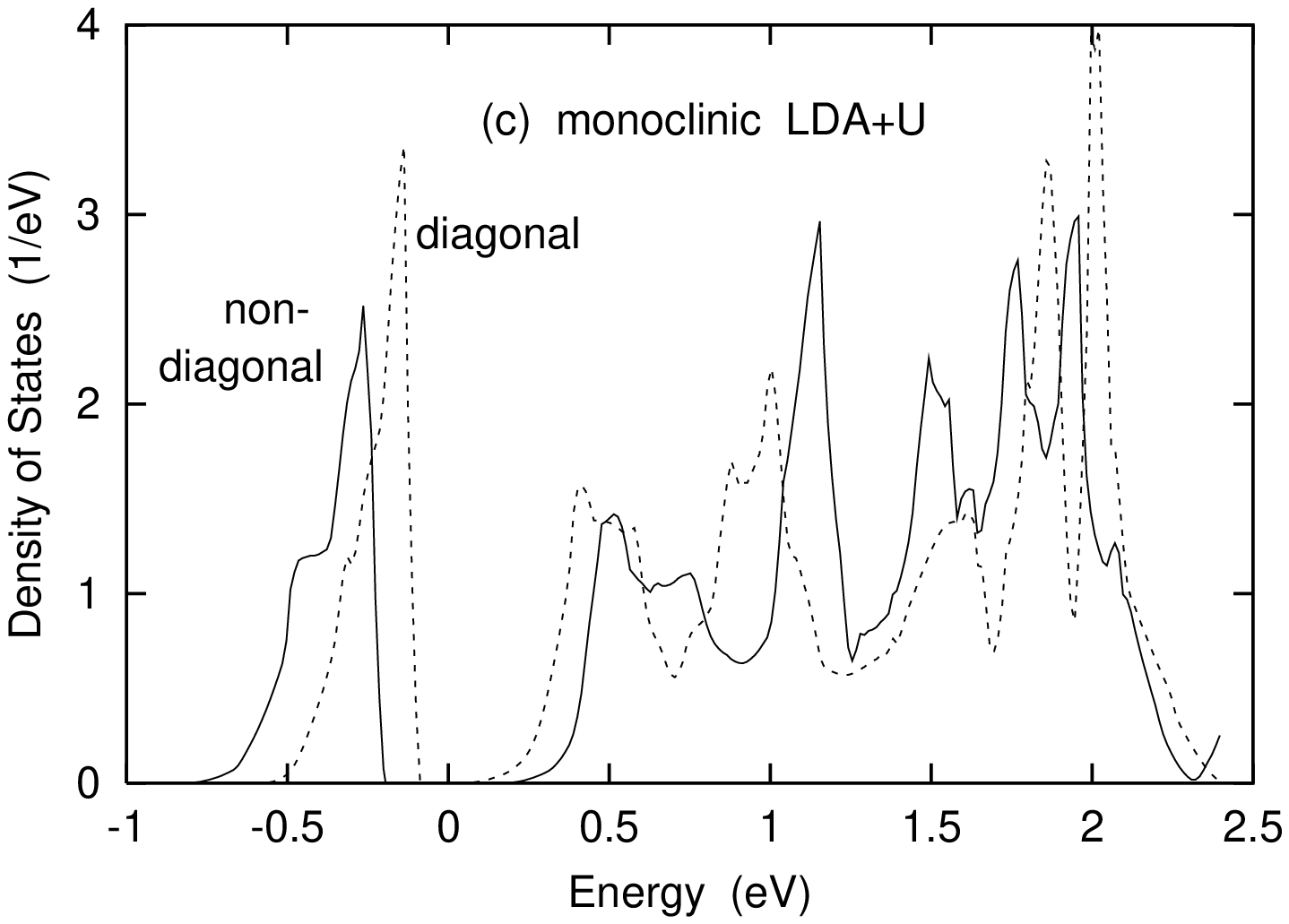}
  \end{center}
  \vskip-4mm
\caption{
(a) VO$_2$ $t_{2g}$ quasi-particle spectra for rutile phase calculated 
within DMFT; (b) $t_{2g}$ partial density of states for monoclinic 
phase calculated within LDA+U; (c) monoclinic density of states 
calculated within non-diagonal (solid curve) and diagonal (dashed curve)
versions of the LDA+U; see text.
Note the different energy scales.
}\end{figure}

\subsection{Insulating Monoclinic Phase}

Turning now to the monoclinic phase we first calculated the $t_{2g}$ 
quasi-particle spectra within the DMFT. Because of the orbital 
polarization caused by the V--V Peierls distortion, correlation effects 
in the $d_{x^2-y^2}$ band are now stronger so that the intensity of the 
lower Hubbard band is enhanced compared to the rutile phase (not shown).
While this trend agrees with the photoemission data, the coherent peak 
near $E_F$ persists. Thus, the single-site DMFT based on a diagonal 
self-energy does not reproduce the insulating nature of the monoclinic 
phase.\cite{heldVO2} This failure is in striking contrast to the results
obtained within the LDA+U and $GW$ methods.\cite{huang,continenza}

Let us discuss first the LDA+U approach. In this scheme different
orbital occupations give rise to orbital dependent potential terms
which shift different $t_{2g}$ bands in opposite directions. Thus,    
orbital polarization leads to strong {\it static} correlation 
effects. Using the same Coulomb parameters as for the rutile 
phase we find that the local density of states calculated via 
the LAPW/LDA+U exhibits a gap of about 0.7~eV (see Fig.~3(b)), in 
agreement with optical data \cite{shin} and previous LDA+U results. 
\cite{huang} The bonding $d_{x^2-y^2}$ bands are now completely 
filled and the $d_{xz,yz}$ and anti-bonding $d_{x^2-y^2}$ bands 
are shifted upwards. Thus, the LDA orbital polarization in the 
monoclinic phase is further increased as a result of static 
correlations. In turn, this reduction of orbital degeneracy 
enhances the trend towards insulating behavior.\cite{koch} 

We emphasize that the LDA+U does not provide a complete description
of correlation effects in the insulating phase. Essentially, it    
amounts to a fully self-consistent, non-local treatment of static 
screening within the dimerized structure. Genuine dynamical effects,
such as the spectral weight transfer from the coherent to the 
incoherent peak as observed in the metallic phase are ignored.   
Neverteless, since static correlations appear to be the origin of 
the excitation gap in VO$_2$, it is instructive to inquire which 
features of the LDA+U contribute to the opening of the gap. 
Of particular interest is the role of the non-diagonal occupation 
matrix $n_{\alpha\beta}$ which is the key input quantity in the 
orbital dependent perturbation potential used in the LDA+U. 
Recent work by Pavarini {\it et al.}~\cite{pavarini} on $3d^1$ 
perovskite materials exhibiting non-diagonal $t_{2g}$ orbital coupling 
caused by octahedral distortions suggests that this mechanism tends 
to suppress orbital fluctuations and to enhance insulating behavior.     
Fig.~3(c) compares the VO$_2$ LDA+U density of states with results 
of an approximate LDA+U treatment in which at each iteration only the 
diagonal elements $n_{\alpha\alpha}$ are retained. The size of the gap 
is seen to be only slightly reduced. In the case of VO$_2$, therefore,
non-diagonal coupling among $t_{2g}$ orbitals is evidently not the 
main reason for the existence of the gap.

To understand the gap formation obtained within the LDA+U, it is 
useful to formally express the $t_{2g}$ self-energy matrix as 
\,$\Sigma(\omega,k)=\Sigma^{\rm HF}(k)+\Delta\Sigma(\omega,k)$,
where $\Sigma^{\rm HF}(k)= H^{\rm LDA+U}(k)-H^{\rm LDA}(k)$ 
is real and accounts for spectral changes 
associated with the LDA+U.\cite{HF-limit} $\Delta\Sigma(\omega,k)$ 
is complex and describes purely dynamical effects. The important 
point is that the LDA+U includes the full momentum variation of 
$\Sigma^{\rm HF}(k)$  
within the true lattice geometry. Thus, in a site representation, 
static screening processes generate finite inter-site elements 
$\Sigma^{\rm HF}_{i\ne j}$ (each element is a matrix in orbital 
space) even if the bare LDA+U perturbation potential is 
site-diagonal. Thus, $\Sigma^{\rm HF}_{ij}(k)$ is more accurate 
than the static limit of the single-site DMFT which neglects the 
$k$ dependence and assumes the impurity environment to be 
isotropic, i.e., $\Sigma^{\rm DMFT}_{ij}(\omega)\sim\delta_{ij}$. 
The results shown in Fig.~3(b,c) suggest that the proper  
evaluation of $\Sigma^{\rm HF}(k)$ within the Brillouin Zone of the 
dimerized structure is the crucial ingredient to an adequate 
description of the insulating behavior in VO$_2$. The LDA+U amounts 
to a self-consistent treatment of $\Sigma^{\rm HF}(k)$ since the 
solution of the Schr\"odinger equation imposes no restrictions on 
how the wave functions adjust to the LDA+U potential. 

In order to go beyond static screening and include dynamical 
correlations in the monoclinic phase a cluster extension of the DMFT 
is most likely required. Such an extension is beyond the scope of 
the present work. A cluster DMFT would include the crucial inter-site 
elements $\Sigma^{\rm DMFT}_{i\ne j}(\omega)$ which arise naturally 
in a cluster representation of the lattice and of the impurity 
Green's functions $G(\omega)$ and $G_0(\omega)$, even for a purely 
on-site Coulomb interaction. Preliminary results for VO$_2$ within a 
cluster DMFT\cite{sascha} show that dynamical screening processes beyond 
the static correlations included in the LDA+U cause a broadening 
of the LDA+U density distribution and a shift of the main spectral 
peak towards the Hubbard bands. Possibly, a multi-site extension 
of the DMFT might also identify the true origin of the metal 
insulator transition in VO$_2$, i.e., whether it is primarily 
caused by the lattice reconstruction or by Coulomb correlations,
or whether these mechanisms mutually enhance each other.      
All one can say at present is that, given the orbital polarization 
induced by the lattice transition, Coulomb correlations have a 
pronounced effect on the quasi-particle spectra. 

As noted above, the $GW$ approach applied to the monoclinic phase 
of VO$_2$ also yields an excitation gap of the correct magnitude. 
\cite{continenza} These calculations utilize a model self-energy 
\cite{gygi} consisting of an approximate short-range contribution
given by the local exchange correlation potential, and a correction   
due to incomplete screening of the Coulomb interaction. Essentially,
in this simplified $GW$ scheme the self-energy correction consists 
of a ``scissor'' operator and additional, non-rigid shifts of energy 
eigenvalues.\cite{gygi}    
Presumably, the reason why this model self-energy yields a gap is 
that $\Sigma(q,\omega)$ is non-local and non-diagonal in site space, 
i.e., it includes the important static correlations in the dimerized 
structure in a similar fashion as the LDA+U. The approximate nature 
of the $GW$ method, in particular, the neglect of electron-electron 
and hole-hole ladder type interactions, affects mainly the remaining 
dynamical corrections caused by the strong local Coulomb energy.
An adequate treatment of these corrections would require going 
beyond the RPA and should lead to a more realistic description 
of the position and width of the Hubbard bands.   

An interesting additional feature observed by Okazaki {\it et al.}
\cite{okazaki} is the temperature dependence of the photoemission 
spectra below the metal insulator transition. Essentially, towards
lower $T$ the excitation gap becomes more clearly defined and the 
main peak near $-1$~eV gets slightly sharper. These changes, however,
are very small on the scale of the main discrepancy still existing 
between the LDA+U or $GW$ results and the experimental data. At 
present it is not clear whether cluster DMFT calculations in the 
monoclinic phase as a function of temperature will be able to 
explain the observed trend or whether an explicit treatment of 
electron-phonon coupling is required.
  
\section{Summary}

The metal insulator transition in VO$_2$ appears to be remarkably
complex and its origins are not yet fully understood. In the present
work we focussed on the important role of two aspects, local Coulomb 
correlations and orbital polarization, in the low and high temperature 
photoemission spectra of VO$_2$. 
Whereas the metallic phase exhibits weak static 
and strong dynamical correlations, the monoclinic phase is dominated 
by static Coulomb correlations. Accordingly, the rutile spectra reveal 
a double-peak structure where the feature close to $E_F$ is identified 
with metallic V $3d$ states and the peak near $-1$\,eV with the lower 
Hubbard band. Since the $t_{2g}$ states in the metallic phase are 
roughly equally occupied orbital polarization is negligible. The 
fundamental difference in the monoclinic phase is the large orbital 
polarization induced by the symmetry breaking due to V--V dimerization. 
The preferential occupation of the $d_{x^2-y^2}$ bonding states implies 
strong static correlations which are the main origin of the excitation 
gap. It would be of great interest to study the additional dynamical
correlation effects in this phase within an extension of the single-site   
DMFT approach.

A. L. likes to thank K. Okazaki, A. Fujimori, and L.H. Tjeng for sharing 
their photoemission data prior to publication. He also thanks 
F. Aryasetiawan, A. Bringer, K. Held, G. Kotliar, A. I. Lichtenstein,
A. Poteryaev, and D. Vollhardt for fruitful discussions.

email: a.liebsch@fz-juelich.de; ishida@chs.nihon-u.ac.jp;
       g.bihlmayer@fz-juelich.de


\begin{thebibliography}{99}

\bibitem{morin} 
   F. J. Morin, Phys. Rev. Lett. {\bf 3}, 34 (1959).

\bibitem{pouget} %
   J. P. Pouget, 
    H. Launois, T. M. Rice, P. Dernier, A. Gossard,  
    G. Villeneuve, and P. Hagenmuller,
   Phys. Rev. B {\bf 10}, 1801 (1977);
   J. P. Pouget, H. Launois, J. P. D'Haenens, P. Merender, and T. M. Rice, 
   Phys. Rev. Lett. {\bf 35}, 873 (1975).

\bibitem{zilbersztejn} %
   A. Zylbersztejn and N. F. Mott,  
   Phys. Rev. B {\bf 11}, 4384 (1975);
 
\bibitem{mott} %
   N. F. Mott, {\it Metal Insulator Transitions} 
   (Taylor and Francis, London, 1990).

\bibitem{rice}   %
   T. M. Rice, H. Launois, and J. P. Pouget,
   Phys. Rev. Lett. {\bf 73}, 3042 (1994).

\bibitem{goodenough} 
   J. B. Goodenough, Phys. Rev. {\bf 117}, 1442 (1960).

\bibitem{caruthers} 
   E. Caruthers and L. Kleinman, 
   Phys. Rev. B {\bf 7}, 3760 (1973).

\bibitem{wentzcovitch} 
   R. M. Wentzcovitch, W. W. Schulz, and P. B. Allen,
   Phys. Rev. Lett. {\bf 72}, 3389 (1994); 
   {\it ibid.} {\bf 73}, 3043 (1994).

\bibitem{eyert} 
   V. Eyert, Ann. Phys.  {\bf 11}, 9 (2002).

\bibitem{huang}
   X. Huang, W. Yang, and U. Eckern,
   cond-mat/9808137.   

\bibitem{korotin}
   M. A. Korotin, N. A. Skorikov, and V. I. Anisimov,
   cond-mat/0301347.   

\bibitem{continenza}
   A. Continenza, S. Massida, and M. Posternak,
   Phys. Rev. B {\bf 60}, 15699 (1999).

\bibitem{laad} %
   M. S. Laad, L. Craco and E. M\"uller-Hartmann,
   cond-mat/0305081.

\bibitem{okazaki}
   K. Okazaki, H. Wadati, A. Fujimori, M. Onoda, Y. Muraoka and Z. Hiroi, 
   Phys. Rev. B. {\bf 69}, 165104 (2004).

\bibitem{tjeng}
   L. H.  Tjeng {\it et al.}, 
   to be published.

\bibitem{DMFT}
   For a review, see:
   A. Georges, G. Kotliar, W. Krauth and M. J. Rozenberg, 
   Rev. Mod. Phys. {\bf 68}, 13 (1996).

\bibitem{LDA+U}
   V. I. Anisimov, J. Zaanen, and O. K. Andersen,
   Phys. Rev. B {\bf 44}, 943 (1991);
   A. I. Liechtenstein, V. I. Anisimov, and J. Zaanen,
   Phys. Rev. B {\bf 52}, R5467 (1995).

\bibitem{GW}
   L. Hedin,  Phys. Rev. A {\bf 139}, 796 (1965).

\bibitem{muraoka}
   Y. Muraoka, Y. Ueda, and Z. Hiroi, 
   J. Phys. Chem. Solids {\bf 63}, 965 (2002).

\bibitem{PES}
   G. A. Sawatzky and D. Post,
   Phys. Rev. B {\bf 20}, 1546 (1979);
   V. M. Bermudez {\it et al.}, {\it ibid.} {\bf 45}, 9266 (1992);
   E. Goering {\it et al.},    {\it ibid.} {\bf 55}, 4225 (1997).

\bibitem{shin}
   S. Shin, 
   S. Suga, M. Taniguchi, M. Fujisawa, H. Kanzaki, A. Fujimori,
   H. Daimon, Y. Ueda, K. Kosuge, and S. Kachi,
   Phys. Rev. B {\bf 41}, 4993 (1990).

\bibitem{maiti}
   K. Maiti, D. D. Sarma, M. J. Rozenberg, I. H. Inoue, H. Makino, O. Goto,
   M. Pedio, and R. Cimino,
   Europhys. Lett. {\bf 55}, 246 (2001).

\bibitem{sekiyama}
   A. Sekiyama, H. Fujiwara, S. Imada, S. Suga, H. Eisaki, S.I. Uchida, 
   K. Takegahara, H. Harima, Y. Saitoh, I. A. Nakrasov, G. Keller, 
   D. E. Kondakov, A. V. Kozhevnikov, Th. Pruschke, K. Held, D. Vollhardt,
   and A. I. Anisimov, Phys. Rev. Lett., to be published. 

\bibitem{liebsch03}
    A. Liebsch, Phys. Rev. Lett. {\bf 90}, 96401 (2003).

\bibitem{perdew}
   J. P. Perdew and Y. Wang, Phys. Rev. B {\bf 45}, 13244 (1992).

\bibitem{UJ}  %
   W. E. Pickett, S. C. Erwin, and E. C. Ethridge,
   Phys. Rev. B {\bf 58}, 1201  (1998). 

\bibitem{jarrell}  %
   M. Jarrell and J. E. Gubernatis,
   Phys. Rep. {\bf 269}, 133 (1996).

\bibitem{pavarini}  %
   E. Pavarini, S. Biermann, A. Poteryaev, A. I. Lichtenstein, A. Georges,
   and O. K. Andersen, Phys. Rev. Lett. {\bf 92}, 176403 (2004).


\bibitem{GWreview}
    W. G. Aulbur, L. J\"onsson and J. W. Wilkins, 
    {\it Quasiparticle Calculations in Solids},  
    Solid State Physics, eds. H. Ehrenreich and F. Spaepen 
    (Academic, San Diego, 2000), Vol. 54, p. 1.

\bibitem{gunnarsson}
   F. Aryasetiawan and  O. Gunnarsson,
   Rep. Progr. Phys. {\bf 61}, 237 (1997).

\bibitem{comment}
   As shown by Huang {\it et al.},\cite{huang} in the rutile phase
   the LDA+U gives a gap only at unrealistically large values of $U$.
   The gap in the insulating phase therefore requires pronounced 
   orbital polarization.

\bibitem{heldVO2}
   This finding agrees with: K. Held {\it et al.}, private communication. 

\bibitem{koch}
   O. Gunnarsson {\it et al.}, Phys. Rev. B {\bf 54}, R11026 (1996);
   E. Koch {\it et al.}, Phys. Rev. B {\bf 60}, 15714 (1999);
   S. Florens {\it et al.}, Phys. Rev. B {\bf 66}, 205102 (2002).  


\bibitem{HF-limit}
   I. Yang, S. Y. Savrasov, and G. Kotliar, 
   cond-mat/ 0209073.

\bibitem{sascha}
   A. Poteryaev and A. I. Lichtenstein, private communication.

\bibitem{gygi}
   F. Gygi and A. Baldereschi, 
   Phys. Rev. Lett. {\bf 62},2160 (1989). 











\end{thebibliography}
\end{document}